# COMPARATIVE ANALYSIS OF SUPERKAMIOKANDE AND SNO SOLAR-NEUTRINO DATA AND THE PHOTOSPHERIC MAGNETIC FIELD

P.A. Sturrock[1]

[1] Center for Space Science and Astrophysics, Varian 302, Stanford University, Stanford, California 94305-4060, U.S.A. (email: sturrock@stanford.edu)

(Received:     )

**Abstract.** We carry out a comparative analysis of Super-Kamiokande, SNO, and photospheric magnetic-field data for the interval in which these datasets overlap. This proves to be the interval of operation of the D$_2$O phase of the SNO experiment. Concerning rotational modulation, we find that the magnetic-field power spectrum shows the strongest peaks at the second and sixth harmonics of the solar synodic rotation frequency [i.e. at $3\nu_{rot}$ and $7\nu_{rot}$]. We find that the restricted Super-Kamiokande dataset has a strong modulation at the second harmonic, as we found to be the case for the complete Super-Kamiokande dataset. The SNO D$_2$O dataset exhibits weak modulation at that frequency, but shows strong modulation in the band corresponding to the sixth harmonic (too high a frequency to be detectable in the Super-Kamiokande dataset, which is available only in 5-day bins, whereas SNO data is available in one-day bins). We estimate the significance level of the correspondence of the Super-Kamiokande second-harmonic peak with the corresponding magnetic-field peak to be 0.0004, and the significance level of the correspondence of the SNO D$_2$O sixth-harmonic peak with the corresponding magnetic-field peak to be 0.009. By estimating the amplitude of the modulation of the solar neutrino flux at the second harmonic from the restricted Super-Kamiokande dataset, we find that the weak power at that frequency in the SNO D$_2$O power spectrum is not particularly surprising. We also examine power spectra in the neighborhood of $9.43\ yr^{-1}$, which is the frequency of a particularly strong modulation in the entire Super-Kamiokande dataset. There is no peak at this frequency in the power spectrum formed from the restricted Super-Kamiokande dataset. It is therefore not surprising that we find (in agreement with the recent analysis by the SNO collaboration) that this peak does not show up in the SNO D$_2$O dataset, either.

# 1. Introduction

Power spectrum analyses of Homestake, GALLEX-GNO, and Super-Kamiokande data have yielded evidence for variability of the solar neutrino flux. Early studies by Haubold (1998), Haubold and Gerth (1990) and Sakurai (1982), and more recent articles summarized and discussed by Caldwell and Sturrock (2005), have now been supplemented by further analyses of GALLEX-GNO (Sturrock, Caldwell and Scargle, 2006) and Super-Kamiokande (Sturrock, Caldwell, Scargle and Wheatland, 2005; Sturrock and Scargle, 2006) data.

Since the Super-Kamiokande experiment (Fukuda et al., 2001, 2002, 2003) and the SNO experiment (SNO Collaboration, 2000) are both sensitive, real-time detectors of solar neutrinos, demonstrable similarity in the time-series patterns in these datasets would support the hypothesis that the solar neutrino flux is variable. The strongest pattern in the Super-Kamiokande data is a modulation with frequency 9.43 $yr^{-1}$ (Sturrock, 2004; Sturrock, Caldwell, Scargle and Wheatland, 2005; Sturrock and Scargle, 2006). The SNO collaboration have recently analyzed their data but found no evidence of this modulation (Aharmin *et al.,* 2005). One of the goals of this article is to study this apparent discrepancy between Super-Kamiokande and SNO.

Concerning this apparent discrepancy, the first point to note is that there is only limited time overlap between these two experiments. The Super-Kamiokande data span the interval 1996.424 to 2001.535, whereas the SNO $D_2O$ and Salt datasets span the intervals 1999.835 to 2001.402 and 2001.566 to 2003.654, respectively. Second, modulation with frequency near 9.43 $yr^{-1}$, if real, is probably attributable to a type of oscillation of which the prototype is one discovered by Rieger et al. (1984). The mechanism of Rieger-type oscillations is still very much a matter of debate [see, for instance, Ballester, Oliver and Carbonnel (2002)]. We have proposed (Sturrock 2004) that they may be interpreted as r-mode oscillations (Papaloizou and Pringle, 1978; Provost, Berthomieu and Rocca, 1981; Saio, 1982). Whatever their mechanism, it is well known that they are typically transient and that the frequency tends to drift (Ballester, Oliver and Carbonnel, 2002; Rybak, Ozguc, Atak and Sozen, 2005). Third, we should note that for a given amplitude of modulation the power derived from power spectrum analysis may vary over a very wide range, so that one experiment may detect the modulation and a similar experiment may miss it (Sturrock, Caldwell,

Scargle, and Wheatland, 2005). In view of these three factors, it is not at all obvious what one should conclude from the fact that a particular oscillation is strong in the complete Super-Kamiokande data but not in the more limited SNO data.

Since – as just mentioned - Rieger-type oscillations are transient and drift in frequency, it is advisable to compare only time-series analyses for the same interval of time. Since there is no time overlap between Super-Kamiokande and the Salt phase of SNO, we analyze only SNO $D_2O$ data and that section of Super-Kamiokande data that covers the same interval of time. We consider it appropriate to supplement these analyses with an analysis of the Sun-center magnetic-field strength for the same interval of time since oscillations of the solar neutrino flux, if real, are probably due to processes involving the solar magnetic field (Akhmedov and Pulido, 2003; Caldwell and Sturrock, 2005; Chauhan and Pulido, 2002; Chauhan and Pulido, 2004; Chauhan, Pulido, and Raghavan, 2005; Miranda et al., 2001).

In Section 2, we carry out a power-spectrum analysis of magnetic-field measurements limited to the interval of operation of the SNO $D_2O$ experiment. Since the measurements are provided in one-day bins, we may extend the power spectrum up to $180\,\mathrm{yr}^{-1}$. We expect there to be peaks related to solar rotation and we search for such peaks. In Section 3, we carry out a power-spectrum analysis of the Super-Kamiokande data, also limited to the interval of operation of the SNO $D_2O$ experiment. Since Super-Kamiokande data has been packaged into five-day bins, this power spectrum extends only to $50\,\mathrm{yr}^{-1}$. We present our power-spectrum analysis of the SNO $D_2O$ data (extending to $180\,\mathrm{yr}^{-1}$) in Section 4.

In Section 5, we compare the power spectra for magnetic-field and Super-Kamiokande data, using one of the statistics recently developed for the comparison and combination of power spectra (Sturrock, Scargle, Walther, and Wheatland, 2005). Section 6 presents a similar comparison for power spectra formed from magnetic-field and SNO $D_2O$ data, and Section 7 presents a similar comparison of the Super-Kamiokande and SNO $D_2O$ power spectra. In Section 8 we present a comparison of all three power spectra (magnetic field, Super-Kamiokande, and SNO $D_2O$).

In Section 9, we attempt to reproduce the power spectrum computed by the SNO Collaboration. There are at least three points of difference between the SNO analysis and our analysis in Section 4, but the most striking difference concerns the calculation of the error in the flux estimates. Our procedure has been to derive the error estimates from the experimental data (as set out in Appendix B). By contrast, the SNO collaboration adopts a quasi-theoretical estimate of the error: their procedure is to calculate the *expected* number of counts in each bin, and to use the square root of this number as a basis for the error estimation. We therefore repeat the power spectrum analysis of Section 4, replacing our procedure for error estimation with the SNO procedure. We discuss our results in Section 10.

## 2. Power-Spectrum of the Photospheric Magnetic Field for the Overlap Interval

We wish to analyze the interval during which the Super-Kamiokande and SNO experiments overlapped. This is effectively the period of operation of the SNO experiment in the $D_2O$ phase, which began on 1999 November 2 and ended on 2001 May 27. The actual duration was 572.2 days, and the live time was 312.9 days. Bin 123 of the Super-Kamiokande 5-day bins started on 1999 November 5, and the last bin (184) ended on 2001 July 15. The SNO Salt phase did not begin until 2001 July 26.

For comparison with the solar neutrino data, we examine the disk-center photospheric magnetic field as measured by the Kitt Peak Solar Observatory. We have formed the average of the (signed) magnetic field over a disk of radius 5 degrees for the time interval 1999 November 2 to 2001 May 27. Since these data are available on a daily basis, with no interruptions, it is sufficient for our purposes to form the Rayleigh power defined by

$$S_R(\nu) = \frac{1}{N} \left| \sum_{k=1}^{N} x_k e^{i2\pi\nu t(k)} \right|^2 , \qquad (1)$$

where $x_k$ is the magnetic field strength normalized to have mean zero and standard deviation unity and N is the total number of data points.

We have found it convenient in solar neutrino research to adopt 1970 January 1 as "Neutrino Day 1," since this predates the commencement of the Homestake Experiment. We then convert the Neutrino Day value to "Neutrino Years" by

$$t(Neutrino\,Years) = 1970 + \frac{t(Neutrino\,Days)}{365.2564} \quad . \tag{2}$$

For comparison, SNO data are listed in terms of "SNO Days" for which Day 1 is 1975 January 1. Hence

$$t(Neutrino\,Days) = t(SNO\,Days) + 1826 \quad . \tag{3}$$

All power spectrum analyses in this article are computed from times measured in "Neutrino Years."

The Rayleigh power of the disk-center (signed) magnetic field strength is shown in Figure 1 in which, since the data are spaced in one-day intervals, we may extend the frequency range to $0 - 180\,\text{yr}^{-1}$ (close to $0.5\,\text{day}^{-1}$). We adopt this frequency range also in our analyses of SNO data, which is also given in one-day bins. The top twenty peaks in this frequency range are listed in Table 1. However, for our analyses of Super-Kamiokande 5-day-binned data, we restrict the frequency range to $0 - 50\,\text{yr}^{-1}$. We therefore show in Figure 2 the Rayleigh power of the disk-center magnetic field strength for this frequency range.

There are two very strong peaks in the band $0 - 180\,\text{yr}^{-1}$: one at $39.96\,\text{yr}^{-1}$ with power 13.04, and the other at $91.53\,\text{yr}^{-1}$ with power 13.85. The equatorial synodic rotation frequency at the photosphere is $13.25\,\text{yr}^{-1}$. However, the magnetic field at disk center may have roots deep in the convection zone, and possibly removed from the equatorial section. Hence it is worth noting that the synodic rotation rate in an equatorial section of the convection zone covers the range $13.1 - 13.9\,\text{yr}^{-1}$, and the synodic rotation rate at latitude $25^\text{o}$ is lower by about $0.3\,\text{yr}^{-1}$ (Schou, Christensen-Dalsgaard, and Thompson, 1992). Hence the peaks at $39.96\,\text{yr}^{-1}$ and $91.53\,\text{yr}^{-1}$ may be identified with the second harmonic and the sixth harmonic, respectively, of the synodic rotation frequency [i.e. $3\nu_{rot}$ and $7\nu_{rot}$]. The peak at $52.83\,\text{yr}^{-1}$ with power 7.22 may be identified as the fifth harmonic of the rotation frequency. In considering the possible modulation of the solar neutrino flux, we have adopted a standard search band of $12.5 - 13.9\,\text{yr}^{-1}$ corresponding to the range of synodic rotation rate in a complete equatorial section solar interior (Sturrock, Caldwell, and Scargle, 2006).

## 3. Power-Spectrum of the Super-Kamiokande Data for the Overlap Interval

In our analysis of the full 5-day-packaged Super-Kamiokande dataset (Sturrock, 2004), we obtained a power-spectrum for the corresponding magnetic-field measurements that is very similar to that shown in Figure 2. We found that one of the three principal peaks in the Super-Kamiokande power spectrum is at frequency at 39.28 $yr^{-1}$, which falls in the band corresponding to the second harmonic of our search band, close to one of the principal peaks in the magnetic-field power spectrum, and we found the time evolution of the modulation in the solar neutrino flux to be similar to the time evolution of the modulation of the disk-center photospheric magnetic field.

We now carry out a power-spectrum analysis of the limited range of Super-Kamiokande (corresponding to the SNO $D_2O$ interval), using a likelihood procedure that takes account of the start and end times and of the error estimates (Sturrock, 2004; Sturrock, Caldwell, Scargle and Wheatland, 2005), which is summarized in Appendix A. The resulting power spectrum is shown in Figure 3. The top twenty peaks for the frequency range $0-50$ $yr^{-1}$ are listed in Table 2. We see that this limited section of Super-Kamiokande data still exhibits a strong peak near 39 $yr^{-1}$ (the actual peak is at 39.38 $yr^{-1}$ with power 9.07). It is interesting to note that what was found to be the principal peak (at 9.43 $yr^{-1}$) for the entire Super-Kamiokande interval is no longer the dominant peak in the power spectrum of the limited dataset. The modulation has shifted in frequency to 9.27 $yr^{-1}$, and the power has dropped to 5.23.

## 4. Power-Spectrum of the SNO $D_2O$ Data for the Overlap Interval

The SNO collaboration has recently released information concerning the live time of their experiment, and the number of solar neutrino events recorded for each day of operation. From these files, we are able to determine, for each day of operation, the duration of the live time, which we denote by d, and the mean live time, which we denote by $t_{ml}$. We may add the mean live time for each day to the date in days to obtain the mean live time as a date. We may then convert this to a date in years, as indicated in Section 2. Using the procedure of Appendix B, we may estimate from the live-time duration d and the number of counts n the following estimate of the flux g (in expected number of counts per day) and the error $\sigma$:

$$g = \frac{n+1}{d} \quad \text{and} \quad \sigma = \frac{(n+1)^{1/2}}{d}. \tag{4}$$

From these data, we may derive a power spectrum by using the Lomb-Scargle procedure (Lomb, 1976; Scargle, 1982), Scargle's extension of that procedure (Scargle, 1989) or, (as we prefer) by the likelihood procedure of Appendix A. The resulting power spectrum is shown in Figure 4 for the frequency band $0-180 \text{ yr}^{-1}$, and in Figure 5 or the frequency band $0-50 \text{ yr}^{-1}$. The top twenty peaks for the band $0-180 \text{ yr}^{-1}$ are shown in Table 3. The biggest peak in the range $0-180 \text{ yr}^{-1}$ is found at $93.83 \text{ yr}^{-1}$ with power 9.45. As is the case for the principal peak in the magnetic-field power spectrum, this falls in a band corresponding to the sixth harmonic of our standard search band $12.5-13.9 \text{ yr}^{-1}$.

## 5. Comparison of the Magnetic-Field and Super-Kamiokande Power Spectra

In Figure 6, we show for comparison purposes the power spectrum of the Super-Kamiokande solar neutrino measurements (shown as positive) and that of the photospheric magnetic field (shown as negative). We see that each power spectrum contains a strong peak near $39 \text{ yr}^{-1}$ - a peak at $39.96 \text{ yr}^{-1}$ with power 13.85 for the magnetic field, and a peak at $39.38 \text{ yr}^{-1}$ with power 9.07 for the Super-Kamiokande data. We also see that each power spectrum also contains a modest peak near $9.3 \text{ yr}^{-1}$ - a peak at $9.30 \text{ yr}^{-1}$ with power 4.45 for the magnetic field, and a peak at $9.27 \text{ yr}^{-1}$ with power 5.23 for the Super-Kamiokande data.

We may carry out a systematic search for pairs of peaks in the two power spectra by using the "joint power statistic," introduced recently (Sturrock, Scargle, Walther and Wheatland, 2005) as a method of looking for correlations in the features of two (or more) power spectra. In comparing two power spectra, we first form

$$X(\nu) = (S_1(\nu)S_2(\nu))^{1/2}, \tag{5}$$

from which the joint power statistic (of second order) is expressible as

$$J = -\ln(2X K_1(2X)), \tag{6}$$

where $K_1$ is the Bessel function of the second kind. We find that the following formula gives a close fit to this statistic:

$$J = \frac{1.943 X^2}{0.650 + X} . \qquad (7)$$

The important point about this statistic is that if the power spectra are distributed exponentially (which is normally the case), then the joint power statistic is also distributed exponentially.

The joint power statistic formed from the Super-Kamiokande and magnetic-field power spectra is shown in Figure 7. The top twenty peaks are listed in Table 4. The leading peaks are seen to be those near $9.3 \text{ yr}^{-1}$ and $39 \text{ yr}^{-1}$. The former is found at $9.28 \text{ yr}^{-1}$ with $J = 7.25$, and the latter is found at $39.65 \text{ yr}^{-1}$ with $J = 8.85$.

It is interesting to obtain an estimate of the significance of the correspondence between the peak at $39.38 \text{ yr}^{-1}$, with power 9.07, in the Super-Kamiokande power spectrum in relation to the principle peak in the magnetic-field power spectrum. Since we are considering only the limited frequency band $0 - 50 \text{ yr}^{-1}$ in analyzing the Super-Kamiokande data, the relevant peak in the magnetic-field power spectrum is that at $39.96 \text{ yr}^{-1}$. We therefore attempt to determine the P-value that estimates the probability of obtaining, by chance, a peak of power 9.07 or more within a band of width $1.16 \text{ yr}^{-1}$ [$2 \times (39.96 - 39.38) \text{ yr}^{-1}$] centered on $39.96 \text{ yr}^{-1}$. We therefore generate 10 000 Monte Carlo simulations of the Super-Kamiokande dataset, extracting a simulated flux value, for each bin, from a normal random distribution centered on the mean value of the flux, with width given by the experimental error estimate for that bin. The result is shown in Figure 8, which displays the result of the simulation in histogram form. We find that only 4 out of 10 000 simulations have power as large as or larger than 9.07 in the prescribed frequency band. This yields a P-Value of 0.0004. Hence this exercise provides good evidence for an association between the Super-Kamiokande power spectrum and the magnetic-field power spectrum.

## 6. Comparison of the Magnetic-Field and SNO Power Spectra

In Figure 9, we show for comparison purposes the power spectrum of the SNO $D_2O$ solar neutrino measurements (shown as positive) and that of the photospheric magnetic field (shown as negative). We see that each power spectrum contains a strong peak near $92 \text{ yr}^{-1}$: the magnetic-field power spectrum (for the SNO $D_2O$ time interval) contains a peak at $91.53 \text{ yr}^{-1}$ with power 13.85, and the SNO $D_2O$ power spectrum contains a peak at $93.83 \text{ yr}^{-1}$ with power 9.45.

The joint power statistic formed from the SNO $D_2O$ and magnetic-field power spectra is shown in Figure 10. The top twenty peaks are listed in Table 5. The leading peak is that at $92.57 \text{ yr}^{-1}$ with $J = 6.76$, but there is also a peak at $39.70 \text{ yr}^{-1}$ with $J = 5.98$.

We now seek an estimate of the significance of the correspondence between the peak at $93.83 \text{ yr}^{-1}$, with power 9.45, in the SNO $D_2O$ power spectrum in relation to the principal peaks in the magnetic-field power spectrum. Since we are now considering the extended frequency band $0 - 180 \text{ yr}^{-1}$, we need to take into account both the peak at $91.53 \text{ yr}^{-1}$ and the peak at $39.96 \text{ yr}^{-1}$ in the magnetic-field power spectrum. We therefore seek to estimate the probability of finding by chance a peak with power 9.45 or more within two bands of width $4.60 \text{ yr}^{-1}$ [$2 \times (93.83 - 91.53) \text{ yr}^{-1}$] centered on $91.53 \text{ yr}^{-1}$ and on $39.96 \text{ yr}^{-1}$.

To this end, we generate 10 000 Monte Carlo simulations of the SNO $D_2O$ dataset by the shuffle procedure, randomly re-assigning measurement data (g and $\sigma$) to the times of the measurements and forming the power spectra as in Section 4. For each simulation, we determine the maximum power in the two frequency bands ($37.66 - 42.26 \text{ yr}^{-1}$ and $89.23 - 93.83 \text{ yr}^{-1}$) taken together. The result is shown in Figure 11, which displays the result of the simulations in histogram form. We find that only 94 out of 10 000 simulations have power as large as or larger than 9.45 in the combined frequency bands. This yields a P-Value of 0.009, providing evidence for an association between the SNO $D_2O$ power spectrum and the magnetic-field power spectrum.

## 7. Comparison of the Super-Kamiokande and SNO Power Spectra

The joint power statistic formed from the Super-Kamiokande and SNO $D_2O$ power spectra for the frequency range $0-50\ \text{yr}^{-1}$ is shown in Figure 11, and the top twenty peaks are listed in Table 6. The two leading peaks are those at $21.29\ \text{yr}^{-1}$ with $J = 6.98$ and at $39.38\ \text{yr}^{-1}$ with $J = 6.95$. The latter gives evidence of the close correspondence between the frequencies of the Super-Kamiokande and SNO $D_2O$ peaks corresponding to the second harmonic of the synodic rotation frequency: they are both found at exactly $39.38\ \text{yr}^{-1}$.

We note that the powers at $39.38\ \text{yr}^{-1}$ are rather different: for Super-Kamiokande, the power is 9.07, whereas for SNO $D_2O$ it is only 2.43. In order to assess the compatibility of these two estimates, we first estimate the depth of modulation of the Super-Kamiokande flux measurements at $39.38\ \text{yr}^{-1}$, which is found to be 10%. We next carry out Monte Carlo simulations of the SNO $D_2O$ measurements for a hypothetical flux that has a 10% depth of modulation at $39.38\ \text{yr}^{-1}$, estimating the value of the count n, for each day, on the basis of Poisson statistics. The results, for 10 000 simulations, is shown in Figure 12. We find that 588 simulations have power as small as or smaller than 2.43. Hence the difference in power is not significant.

We comment also on the fact, noted by the SNO Collaboration (Aharmin et al., 2005), that the power spectrum formed from the SNO $D_2O$ dataset does not show a peak at $9.43\ \text{yr}^{-1}$, whereas the power spectrum formed from the complete Super-Kamiokande five-day dataset shows a very strong peak at that frequency. Our most recent analysis, which takes account of the asymmetry in the error estimates, yields a power of 13.24 at that frequency (Sturrock and Scargle, 2006). In carrying out 10 000 Monte Carlo simulations, we found only 11 simulations that had that strong a power or stronger in a frequency band chosen to be $1-36\ \text{yr}^{-1}$. The important point to note is that the power spectrum formed from that part of the Super-Kamiokande dataset that falls within the SNO $D_2O$ time interval does not show a peak at $9.43\ \text{yr}^{-1}$. As we see from Table 2, the peak nearest this frequency is found at $9.27\ \text{yr}^{-1}$ and has power only 5.23. Hence the fact that SNO $D_2O$ data do not reveal a peak at $9.43\ \text{yr}^{-1}$ is quite compatible with what we find in the Super-Kamiokande power spectrum for the same time interval.

As we have suggested elsewhere (Sturrock and Scargle, 2006), it would probably be more appropriate, and may avoid future confusion, if one were to use the term "quasi-periodic" in referring to modulations that appear in analyses of solar neutrino data. We have proposed that the prominent modulation at $9.43 \text{ yr}^{-1}$ in the entire Super-Kamiokande dataset is similar to the well-known Rieger-type modulations (Rieger et al., 1984; Ballester, Oliver, and Carbonel, 2002; Rybak et al., 2005), which we suggest have their origin in r-mode oscillations (Sturrock, 2004). Rieger-type oscillations are known to be transient phenomena, which may therefore be referred to as "quasi-periodic."

## 8. Comparison of Magnetic-Field, Super-Kamiokande, and SNO Power Spectra

It is interesting to search for common features in the power spectra derived in Sections 2, 3 and 4 from magnetic-field, Super-Kamiokande, and SNO data for the overlap time interval (the SNO $D_2O$ interval). A convenient procedure is that of constructing the joint power statistic (Sturrock, Scargle, Walther, and Wheatland, 2005). We first form the geometric mean

$$X(\nu) = \left(S_1(\nu)S_2(\nu)S_3(\nu)\right)^{1/3}. \quad (8)$$

Then the joint-power statistic is given, to sufficient accuracy, by

$$J = \frac{2.916 X^2}{1.022 + X}. \quad (9)$$

This statistic is shown, for the frequency range $0 - 50 \text{ yr}^{-1}$, in Figure 13, and the top twenty peaks are listed in Table 7.

We see that the most prominent feature is the peak at $39.54 \text{ yr}^{-1}$ with $J = 8.55$, corresponding to the second harmonic of the synodic rotation frequency. The second most prominent feature is the peak at $9.24 \text{ yr}^{-1}$ with $J = 6.07$, which we may identify with the oscillation (presumed to be an r-mode oscillation) which is prominent (at the nearby frequency $9.43 \text{ yr}^{-1}$) in the power spectrum formed from the complete Super-Kamiokande dataset. The third feature, at $27.17 \text{ yr}^{-1}$ with $J = 3.92$, may be the first harmonic of the synodic rotation frequency in the convection zone.

## 9. The SNO Analysis

The results of our power-spectrum analysis differ significantly from those of the SNO collaboration (Aharmin et al. 2005). Since we analyze the same data, the difference must be due to differences in the analyses. The most striking difference concerns the calculation of the error in the flux estimates. Our procedure has been to derive the error estimates from the experimental data, as set out in Appendix B. By contrast, the SNO collaboration adopts a quasi-theoretical estimate of the error: their procedure is to calculate the expected number of counts in each bin, and to use the square root of this number as a basis for the error estimation.

In terms of the notation of Appendix B, our understanding is that the SNO collaboration estimate the flux and error for each bin from

$$g_{SNO} = \frac{n}{d} \quad \text{and} \quad \sigma_{SNO} = \frac{(fd)^{1/2}}{d}, \qquad (10)$$

where f is the "mean event rate" given by

$$f = \frac{\sum n_r}{\sum d_r}. \qquad (11)$$

Our estimates of the errors are very different from the SNO estimates. Our estimates have a mean of 5.49 and a standard deviation of 9.50. By contrast, the SNO estimates have a mean of 8.02 and a standard deviation of 2.97. It should therefore not be surprising if the resulting power spectra are also very different.

The SNO collaboration also adopts the following procedure: *Any bin in which fewer than five counts would be expected based upon the bin's livetime and the mean event rate was combined with the following bin(s) so that the expected number of events in all bins was greater than five.* We find that 142 bins out of the total of 953 bins of $D_2O$ data (15%) would need to be merged with other bins. We choose not to merge bins for this or any other reason.

Another difference between our analysis and the SNO analysis is that we use a likelihood procedure whereas the SNO Collaboration uses the Scargle (1989) version of the Lomb-Scargle procedure (Lomb, 1976; Scargle, 1982). In the notation of Appendix A, the Scargle procedure uses the mean value of $g_r$ as an estimate of $G_0$, whereas we use the maximum-likelihood value. In the present analysis, the difference is significant: the mean value of the flux is 12.72, but the maximum-likelihood value is 9.49. Hence we cannot expect the Scargle procedure to yield the same power spectrum as the likelihood procedure. Whereas the likelihood procedure gives a peak of power 9.45 at 93.83 $yr^{-1}$, the Scargle procedure gives a peak of only 6.13 at that frequency.

In this section, we investigate only the consequence of estimating errors on the basis of the expected number of counts rather than the actual number of counts in each bin. We have repeated the power-spectrum analysis of Section 4, using the flux and error estimates given by (10) and (11) rather than (B.7). If we were to find that the results are close to those given in Section 4, we could conclude that it makes no difference whether we use an error estimate based on the expected number of counts, or an estimate based on the actual number of counts. On the other hand, if there is a significant difference between the results, our conclusion will be that one should use the actual number of counts rather than the expected number.

The results of this calculation are shown in Figures 14 and 15 for the frequency ranges and $0-50$ $yr^{-1}$ for comparison with Figures 4 and 5, respectively. The top twenty peaks for the range $0-180$ $yr^{-1}$ are shown in Table 8. In order to facilitate comparison with the SNO article, we also show the power spectrum with frequency measured in cycles per day rather than cycles per year. The result, for the range $0-0.5$ $day^{-1}$, is shown in Figure 16.

## 10. Discussion

We now review the results of our analyses. In our analysis of the photospheric magnetic field in Section 2, we found two very strong peaks in the band $0-180\,yr^{-1}$: one at $39.96\,yr^{-1}$ with power 13.04, and the other at $91.53\,yr^{-1}$ with power 13.85. In previous articles concerned with the possible modulation of the solar neutrino flux, we have adopted a standard search band of $12.5-13.9\,yr^{-1}$ corresponding to the range of synodic rotation rate in a complete equatorial section (radiative zone and convection zone) of the solar interior (Schou, Christenen-Dalsgaard, and Thompson, 1992). Hence the peaks at $39.96\,\text{yr}^{-1}$ and at $91.53\,\text{yr}^{-1}$ may be identified with the second harmonic and the sixth harmonic, respectively, of the synodic rotation frequency [i.e. $3\nu_{rot}$ and $7\nu_{rot}$]. The peak at $52.83\,\text{yr}^{-1}$ with power 7.22 may be identified as the fifth harmonic of the rotation frequency. The inferred rotation frequencies ($13.32\,\text{yr}^{-1}, 13.08\,\text{yr}^{-1}$, and $13.21\,\text{yr}^{-1}$, respectively) all fall within the band of equatorial rotation frequencies of the convection zone (approximately $13.1-13.9\,\text{yr}^{-1}$; Schou, Christenen-Dalsgaard, and Thompson, 1992). These results are reasonable, since it is to be expected that the photospheric magnetic field will display rotation patterns characteristic of the convection zone.

In our power-spectrum analysis of Super-Kamiokande data for the overlap time interval in Section 3, we noted that, by comparison with our power-spectrum analyses of the entire Super-Kamiokande dataset (Sturrock, 2004; Sturrock, Caldwell, Scargle, and Wheatland, 2005; Sturrock and Scargle, 2006), we still find a strong peak ($39.38\,\text{yr}^{-1}$ with power 9.07) corresponding to the second harmonic of the rotation frequency. However, it is important to note that what was found to be the principal peak (at $9.43\,\text{yr}^{-1}$) for the entire Super-Kamiokande interval is no longer the dominant peak in the power spectrum of the limited dataset. The modulation has shifted in frequency to $9.27\,\text{yr}^{-1}$, and the power has dropped to 5.23. If the Super-Kamiokande power spectrum does not show a strong peak at $9.43\,\text{yr}^{-1}$, there is little reason to expect that the SNO power spectrum will do so. Hence it should come as no surprise that the SNO power spectrum does not display a prominent peak at $9.43\,\text{yr}^{-1}$. Such a drift in frequency and change in power are quite consistent with our proposed interpretation of the oscillation as a Rieger-type oscillation (Sturrock, 2004). It is interesting to note from Table 1 that the magnetic-field power spectrum shows a peak at $9.30\,\text{yr}^{-1}$ with power 4.45, supporting an interpretation of that peak as attributable to an internal oscillation.

Our analysis of SNO $D_2O$ data in Section 4 shows that the most prominent peak in the power spectrum is found at $93.83\,\text{yr}^{-1}$ with power 9.45. This may be attributed to the sixth harmonic of the rotation frequency, in the same frequency band as the principal peak in the magnetic-field power spectrum. Interestingly enough, we also find a peak at $39.38\,\text{yr}^{-1}$ (but with much lower power, 2.43) which is exactly the same frequency as the second most prominent peak in the Super-Kamiokande power spectrum for this time interval.

In our comparison of the magnetic-field and Super-Kamiokande power spectra in Section 5, we carried out Monte Carlo simulations to evaluate the apparent correspondence between the Super-Kamiokande peak at $39.38 \, yr^{-1}$ with power 9.07 and the strongest magnetic field peak in the band $0-50 \, yr^{-1}$, which occurs at $39.96 \, yr^{-1}$. We found that only 4 out of 10 000 simulations have peaks that are as close to the magnetic-field peak or closer, and have power as large as or larger than 9.07, yielding a P-Value of 0.0004 in favor of an association between the two features.

Section 6 presents our comparison of the magnetic-field and the SNO $D_2O$ power spectra. The strongest peak in the SNO $D_2O$ power spectrum is at $93.83 \, yr^{-1}$ with power 9.45. This is in the band of frequencies corresponding to the sixth harmonic of the rotation frequency, which also contains the strongest peak in the magnetic-field power spectrum. We estimate, from Monte Carlo simulations, that the association between the leading peak in the SNO $D_2O$ power spectrum and the two leading peaks of the magnetic-field power spectrum would occur by chance with probability 0.009.

In Section 7, we compare the Super-Kamiokande and SNO $D_2O$ power spectra, by forming the joint power statistic. We find that this statistic has two notable peaks, one at $21.29 \, yr^{-1}$ with $J = 6.98$, and the other at $39.38 \, yr^{-1}$ with power $J = 6.95$. We noted that the second peak gives evidence of the close correspondence between the frequencies of the Super-Kamiokande and SNO $D_2O$ peaks corresponding to the second harmonic of the synodic rotation frequency.

Section 8 presents the joint power statistic formed from the three power spectra derived from the magnetic-field, Super-Kamiokande, and SNO $D_2O$ datasets. The most prominent feature is the peak at $39.54 \, yr^{-1}$ with $J = 8.55$, corresponding to the second harmonic of the synodic rotation frequency. The second most prominent feature is the peak at $9.24 \, yr^{-1}$ with $J = 6.07$. It seems likely that this may be identified with the oscillation (presumed to be an r-mode oscillation) that is prominent (at the nearby frequency $9.43 \, yr^{-1}$) in the power spectrum formed from the complete Super-Kamiokande dataset. The third feature, at 27.17 yr$^{-1}$ with $J = 3.92$, may be the first harmonic of the synodic rotation frequency in the convection zone.

Concerning our attempt to reproduce the SNO power spectrum analysis in Section 9, we see that Figures 14 and 15 differ significantly from Figures 4 and 5. However, Figure 16 is similar to the corresponding figure in the SNO article (Aharmin et al., 2005). For instance, the strongest peak in Table 8 is at frequency 149.13 $yr^{-1}$ or 0.4083 $day^{-1}$ with power 7.96. The strongest peak in the SNO power spectrum is at this frequency, with power approximately 7.3. On the other hand, we see from Table 3 that the strongest peak in our power spectrum derived from SNO D$_2$O data is at 93.83 $yr^{-1}$ with power 9.45. This peak is the fifth strongest in Table 8, and has power 5.26. Furthermore, we have noted in Section 6 that our power spectrum contains a peak close to the lower of the two principal peaks in the magnetic-field power spectrum, namely a peak at 39.38 $yr^{-1}$ with power 2.43. The nearest peak in our SNO-type power spectrum is at 39.34 $yr^{-1}$ with power only 0.69, which is completely insignificant.

We conclude that a power-spectrum analysis of the SNO D$_2$O dataset that (a) uses a likelihood procedure, (b) employs error estimates derived from the experimental data, and (c) retains the separation between bins, shows points of correspondence with power spectra derived from magnetic-field data and Super-Kamiokande data for the same time interval.

I wish to thank the SNO and Super-Kamiokande collaborations for making their data available, and to thank David Caldwell, Alexander Kosovichev, Joao Pulido, Jeffrey Scargle, Guenther Walter, Michael Wheatland, and Steven Yellin for their continued interest and support.

**Appendix A. Likelihood Analyses**

The likelihood that flux measurements g$_r$ may be fitted to a functional form G$_r$ is given, ignoring some multiplying constants, by

$$Lik = \prod_{r=1}^{R} \exp\left[-\frac{1}{2}\frac{(g_r - G_r)^2}{\sigma_r^2}\right]. \tag{A.1}$$

Hence the log-likelihood is given by

$$L = -\frac{1}{2}\sum_{r=1}^{R}\frac{(g_r - G_r)^2}{\sigma_r^2}. \tag{A.2}$$

We first determine a constant value $G_0$ that maximizes the log-likelihood, giving the value $L_0$.

If (as in the SNO calculations), the measurements are assigned to discrete times $t_r$, then we consider the functional form

$$G_r = G_0 + Ae^{i2\pi \nu t_r} + A^* e^{-i2\pi \nu t_r}. \tag{A.3}$$

For each value of the frequency $\nu$, we adjust the complex amplitude A to maximize the log-likelihood. We then define the power S by

$$S = L - L_0. \tag{A.4}$$

Note that it possible, in this procedure, to re-adjust the term $G_0$ for each frequency. This is the "floating offset" procedure (Koshio 2003; Sturrock *et al.* 2005). However, this procedure is ill determined at zero frequency so it may yield unreasonably large values of the power for small frequencies. This modification usually makes little difference to the power estimates, so we do not use it in the present calculations.

If (as in the Super-Kamiokande calculations), the flux estimates are derived by integrating counts over a finite time interval, we need to modify Equation (A.3) as follows:

$$G_r = G_0 + \frac{1}{t_{er} - t_{sr}}\int_{t_{sr}}^{t_{er}} dt\left[Ae^{i2\pi \nu t} + A^* e^{-i2\pi \nu t}\right]. \tag{A.5}$$

**Appendix B. SNO Flux and Error Estimates**

Given the live-time duration d and the number of counts n for each day, we need to obtain an estimate of the flux g and an estimate of the error $\sigma$. If the flux and error are to be measured in counts per day, the expected number of counts m is given by $m = gd$.

For given m, the probability distribution function for n is the Poisson distribution:

$$P(n\,|\,m) = F_P(n,m) \equiv \frac{m^n}{n!} e^{-m}. \tag{B.1}$$

We need $P(m\,|\,n)$, which is related to $P(n\,|\,m)$ by

$$P(m\,|\,n) = \frac{P(n\,|\,m)}{\sum_{m'} P(n\,|\,m') P(m'\,|-)} P(m\,|-). \tag{B.2}$$

If we adopt a uniform prior distribution, then

$$P(m\,|\,n) = \frac{P(n\,|\,m)}{\sum_{m'} P(n\,|\,m')}, \tag{B.3}$$

from which it follows that

$$P(m\,|\,n) = F_P(n,m) \equiv \frac{m^n}{n!} e^{-m}, \tag{B.4}$$

so that

$$\langle m \rangle = \int_0^\infty dm \, \frac{m^{n+1}}{n!} e^{-m} = n+1 \tag{B.5}$$

and

$$\langle m^2 \rangle = \int_0^\infty dm \, \frac{m^{n+2}}{n!} e^{-m} = (n+1)(n+2). \tag{B.6}$$

Hence, given the count n and the duration d, the flux estimate and the error estimate of the flux are given by

$$g = \frac{n+1}{d} \quad \text{and} \quad \sigma = \frac{(n+1)^{1/2}}{d}. \tag{B.7}$$

is completely insignificant.

TABLE 1

Top twenty peaks over the frequency range $0$ to $180\,yr^{-1}$ in the power spectrum formed from the magnetic field dataset.

| Order | Frequency (yr$^{-1}$) | Power | Order | Frequency (yr$^{-1}$) | Power |
|---|---|---|---|---|---|
| 1 | 91.53 | 13.85 | 11 | 9.3 | 4.45 |
| 2 | 39.96 | 13.04 | 12 | 106.92 | 4.4 |
| 3 | 92.64 | 9.52 | 13 | 4.03 | 3.98 |
| 4 | 52.83 | 7.22 | 14 | 118.32 | 3.61 |
| 5 | 38.74 | 5.73 | 15 | 104.71 | 3.41 |
| 6 | 6.67 | 5.03 | 16 | 94.45 | 3.19 |
| 7 | 13.87 | 4.85 | 17 | 1.47 | 3.16 |
| 8 | 2.68 | 4.81 | 18 | 90.56 | 3.13 |
| 9 | 103.49 | 4.71 | 19 | 171.2 | 3.1 |
| 10 | 119.54 | 4.47 | 20 | 51.62 | 2.95 |

TABLE 2

Top twenty peaks over the frequency range $0$ to $50\,yr^{-1}$ in the power spectrum formed from the Limited Super-Kamiokande dataset.

| Order | Frequency (yr$^{-1}$) | Power | Order | Frequency (yr$^{-1}$) | Power |
|---|---|---|---|---|---|
| 1 | 48.58 | 9.77 | 11 | 3.20 | 3.33 |
| 2 | 39.38 | 9.07 | 12 | 18.08 | 3.31 |
| 3 | 43.69 | 6.46 | 13 | 22.77 | 3.01 |
| 4 | 21.27 | 5.51 | 14 | 46.63 | 2.95 |
| 5 | 9.27 | 5.23 | 15 | 12.89 | 2.86 |
| 6 | 31.43 | 4.36 | 16 | 12.43 | 2.77 |
| 7 | 26.90 | 4.07 | 17 | 30.58 | 2.77 |
| 8 | 11.21 | 4.05 | 18 | 20.37 | 2.76 |
| 9 | 41.44 | 3.80 | 19 | 33.70 | 2.54 |
| 10 | 5.77 | 3.52 | 20 | 34.68 | 2.45 |

TABLE 3

Top twenty peaks over the frequency range $0$ to $180\,yr^{-1}$ in the power spectrum formed from the SNO D$_2$O dataset.

| Order | Frequency (yr$^{-1}$) | Power | Order | Frequency (yr$^{-1}$) | Power |
|---|---|---|---|---|---|
| 1 | 93.83 | 9.45 | 11 | 127.11 | 4.73 |
| 2 | 97.86 | 6.48 | 12 | 84.94 | 4.54 |
| 3 | 101.10 | 6.08 | 13 | 6.97 | 4.36 |
| 4 | 150.06 | 6.06 | 14 | 60.13 | 4.06 |
| 5 | 150.90 | 5.84 | 15 | 113.97 | 4.05 |
| 6 | 159.06 | 5.55 | 16 | 21.47 | 4.02 |
| 7 | 104.47 | 5.51 | 17 | 161.27 | 4.02 |
| 8 | 149.18 | 5.39 | 18 | 72.77 | 3.95 |
| 9 | 69.24 | 5.18 | 19 | 108.47 | 3.91 |
| 10 | 124.25 | 4.86 | 20 | 30.90 | 3.90 |

TABLE 4

Top twenty peaks over the frequency range $0$ to $50\,yr^{-1}$ for the joint power statistic formed from the magnetic-field and Super-Kamiokande power spectra.

| Order | Frequency (yr$^{-1}$) | JPS | Order | Frequency (yr$^{-1}$) | JPS |
|---|---|---|---|---|---|
| 1 | 39.65 | 8.85 | 11 | 4.18 | 2.59 |
| 2 | 9.28 | 7.25 | 12 | 43.66 | 2.09 |
| 3 | 48.58 | 6.21 | 13 | 14.96 | 1.98 |
| 4 | 39.03 | 5.65 | 14 | 17.42 | 1.93 |
| 5 | 41.38 | 3.55 | 15 | 20.22 | 1.91 |
| 6 | 27.08 | 3.53 | 16 | 11.39 | 1.91 |
| 7 | 2.46 | 3.51 | 17 | 12.3 | 1.89 |
| 8 | 3.08 | 3.48 | 18 | 13.11 | 1.84 |
| 9 | 1.58 | 3.02 | 19 | 29.1 | 1.79 |
| 10 | 13.67 | 2.88 | 20 | 5.66 | 1.73 |

TABLE 5

Top twenty peaks over the frequency range $0$ to $180\,yr^{-1}$ for the joint power statistic formed from the magnetic-field and SNO $D_2O$ power spectra.

| Order | Frequency (yr$^{-1}$) | JPS | Order | Frequency (yr$^{-1}$) | JPS |
|---|---|---|---|---|---|
| 1 | 92.57 | 6.76 | 11 | 9.19 | 3.62 |
| 2 | 104.56 | 6.58 | 12 | 127.16 | 3.56 |
| 3 | 6.84 | 6.55 | 13 | 150.88 | 3.45 |
| 4 | 93.72 | 6.33 | 14 | 132.13 | 3.36 |
| 5 | 39.70 | 5.98 | 15 | 118.56 | 3.34 |
| 6 | 101.04 | 5.06 | 16 | 107.57 | 3.30 |
| 7 | 119.61 | 4.94 | 17 | 38.98 | 3.30 |
| 8 | 103.47 | 4.84 | 18 | 144.36 | 3.26 |
| 9 | 91.16 | 4.40 | 19 | 85.02 | 3.22 |
| 10 | 27.34 | 4.01 | 20 | 102.60 | 3.19 |

TABLE 6

Top twenty peaks over the frequency range $0$ to $50\,yr^{-1}$ for the joint power statistic formed from the Super-Kamiokande and SNO $D_2O$ power spectra.

| Order | Frequency (yr$^{-1}$) | JPS | Order | Frequency (yr$^{-1}$) | JPS |
|---|---|---|---|---|---|
| 1 | 21.29 | 6.98 | 11 | 41.64 | 2.80 |
| 2 | 39.38 | 6.95 | 12 | 27.12 | 2.64 |
| 3 | 11.19 | 4.31 | 13 | 45.77 | 2.62 |
| 4 | 46.65 | 4.19 | 14 | 26.77 | 2.49 |
| 5 | 30.70 | 3.71 | 15 | 3.16 | 2.44 |
| 6 | 9.20 | 3.51 | 16 | 0.64 | 2.25 |
| 7 | 48.32 | 3.50 | 17 | 33.59 | 2.16 |
| 8 | 31.27 | 3.35 | 18 | 43.50 | 1.88 |
| 9 | 20.36 | 2.87 | 19 | 17.34 | 1.86 |
| 10 | 27.60 | 2.81 | 20 | 21.91 | 1.85 |

TABLE 7

Top twenty peaks over the frequency range 0 to $50\,yr^{-1}$ for the joint power statistic formed from the magnetic-field, Super-Kamiokande and SNO D$_2$O power spectra.

| Order | Frequency ($yr^{-1}$) | JPS | Order | Frequency ($yr^{-1}$) | JPS |
|---|---|---|---|---|---|
| 1 | 39.54 | 8.55 | 11 | 17.36 | 2.59 |
| 2 | 9.24 | 6.07 | 12 | 21.11 | 2.50 |
| 3 | 27.17 | 3.92 | 13 | 2.48 | 2.48 |
| 4 | 48.47 | 3.78 | 14 | 6.78 | 2.44 |
| 5 | 3.09 | 3.44 | 15 | 1.62 | 2.34 |
| 6 | 41.62 | 3.30 | 16 | 13.70 | 1.96 |
| 7 | 27.45 | 3.04 | 17 | 14.95 | 1.94 |
| 8 | 11.39 | 2.98 | 18 | 41.21 | 1.89 |
| 9 | 20.21 | 2.76 | 19 | 12.15 | 1.89 |
| 10 | 30.74 | 2.61 | 20 | 36.88 | 1.86 |

TABLE 8

Top twenty peaks over the frequency range 0 to $180\,yr^{-1}$ in the power spectrum formed by the SNO procedure from the SNO D$_2$O dataset.

| Order | Freq ($yr^{-1}$) | Freq ($day^{-1}$) | Power | Order | Freq ($yr^{-1}$) | Freq ($day^{-1}$) | Power |
|---|---|---|---|---|---|---|---|
| 1 | 149.13 | 0.4083 | 7.96 | 11 | 64.76 | 0.1773 | 4.51 |
| 2 | 104.52 | 0.2862 | 5.91 | 12 | 69.12 | 0.1892 | 4.48 |
| 3 | 72.77 | 0.1992 | 5.55 | 13 | 150.86 | 0.4130 | 4.28 |
| 4 | 101.17 | 0.2770 | 5.48 | 14 | 21.13 | 0.0578 | 4.12 |
| 5 | 93.79 | 0.2568 | 5.26 | 15 | 11.73 | 0.0321 | 4.02 |
| 6 | 70.34 | 0.1926 | 5.04 | 16 | 161.33 | 0.4417 | 3.88 |
| 7 | 150.03 | 0.4108 | 4.98 | 17 | 97.83 | 0.2678 | 3.80 |
| 8 | 132.95 | 0.3640 | 4.67 | 18 | 108.58 | 0.2973 | 3.63 |
| 9 | 45.75 | 0.1253 | 4.65 | 19 | 187.51 | 0.5134 | 3.45 |
| 10 | 124.29 | 0.3403 | 4.58 | 20 | 188.41 | 0.5158 | 3.34 |

FIGURES

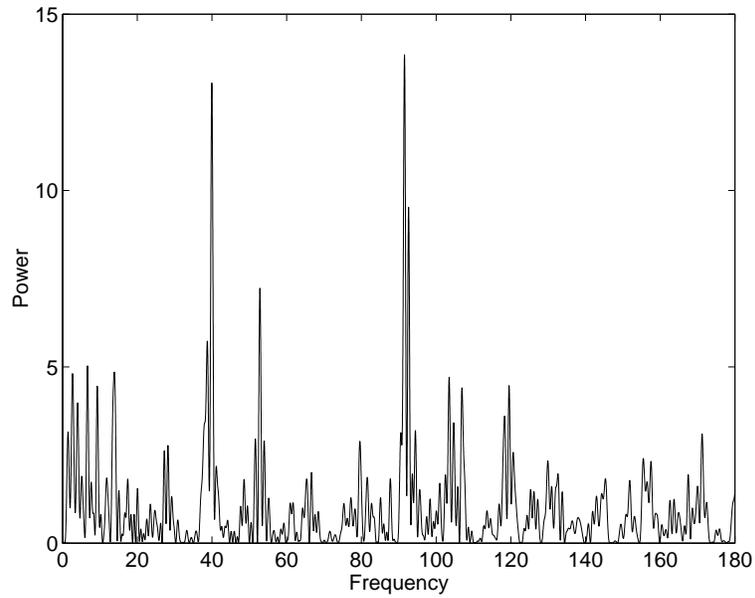

Figure 1. The Rayleigh power formed from the limited magnetic-field dataset for the frequency range $0-180\ yr^{-1}$.

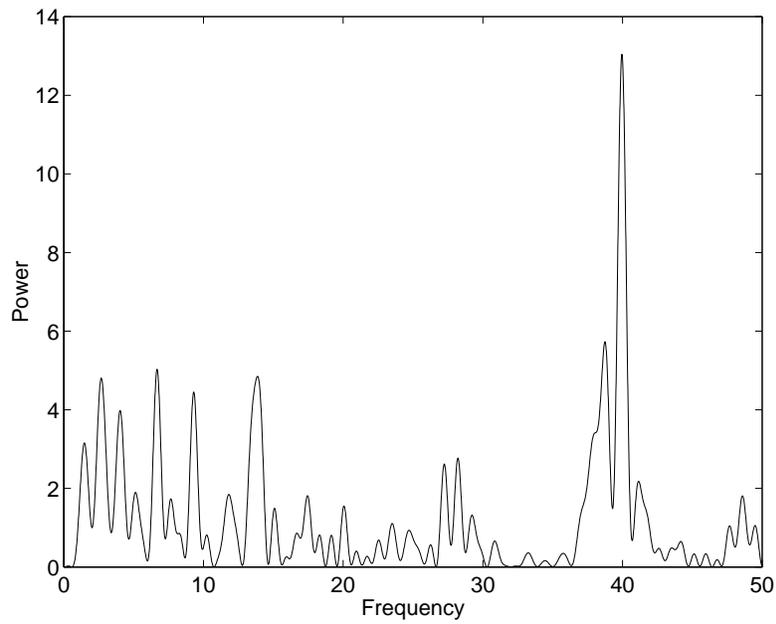

Figure 2. The Rayleigh power formed from the limited magnetic-field dataset for the frequency range $0-50\ yr^{-1}$.

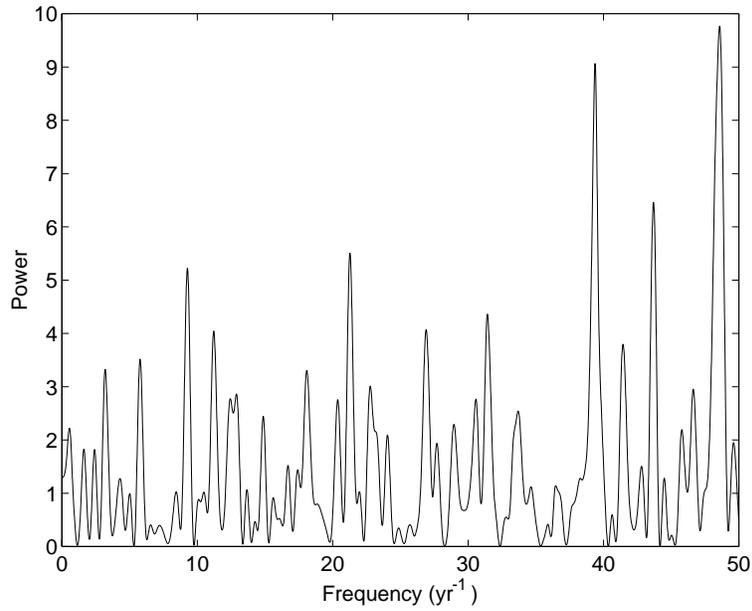

Figure 3. Power spectrum of the Limited Super-Kamiokande dataset, computed by a likelihood procedure for the frequency range $0-50\ yr^{-1}$.

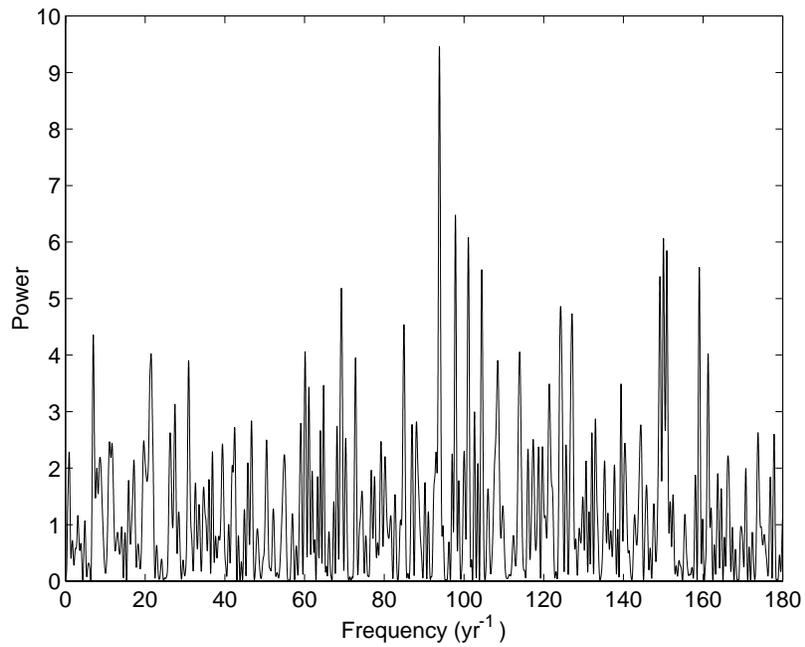

Figure 4. Power spectrum of the $D_2O$ section of the SNO dataset, computed by a likelihood procedure for the frequency range $0-180\ yr^{-1}$.

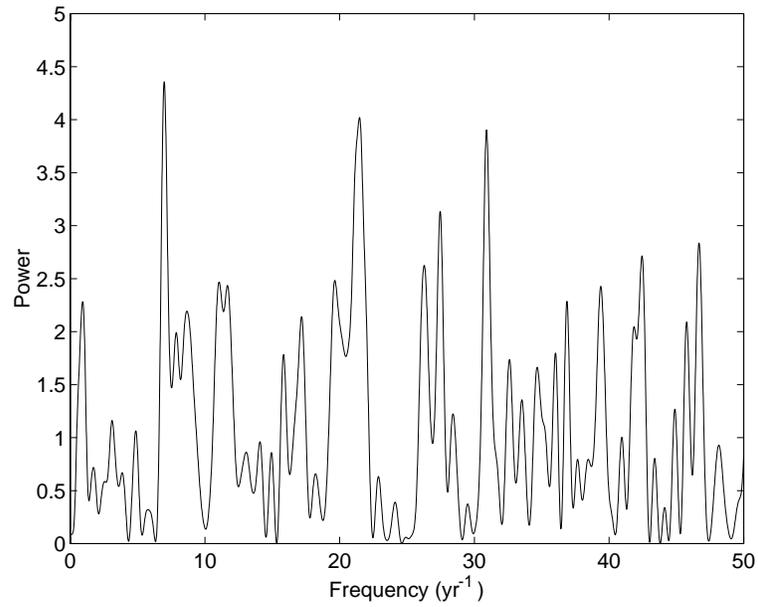

Figure 5. Power spectrum of the $D_2O$ section of the SNO dataset, computed by a likelihood procedure for the frequency range $0-50$ $yr^{-1}$.

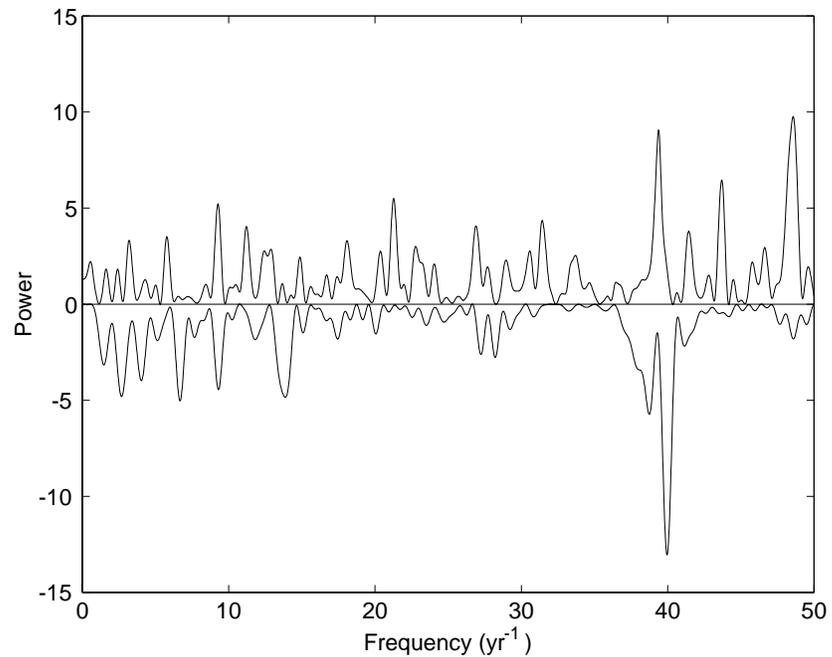

Figure 6. Display of the Super-Kamiokande power spectrum (shown positive) and the magnetic-field power spectrum (shown negative) during the SNO $D_2O$ interval.

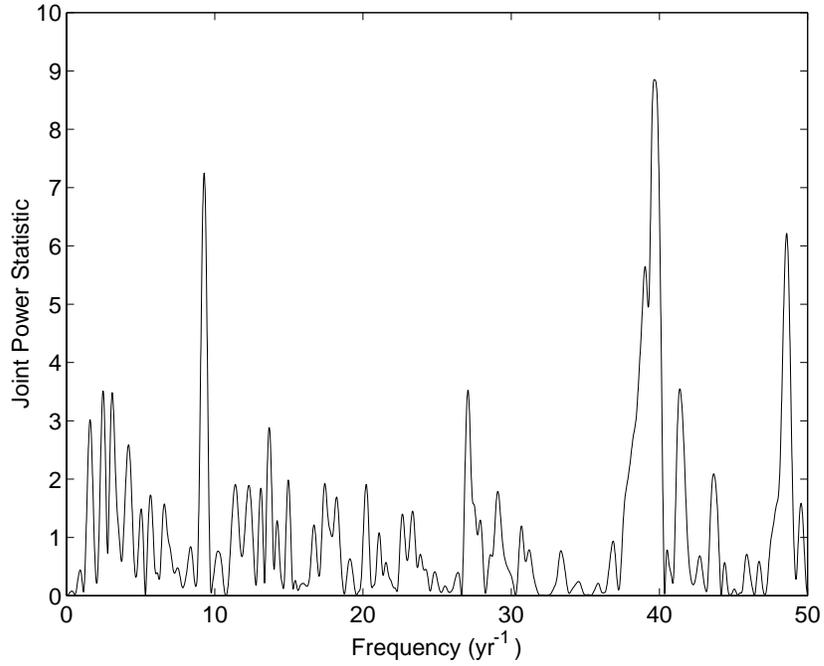

Figure 7. The joint power statistic formed from the magnetic field and Super-Kamiokande power spectra.

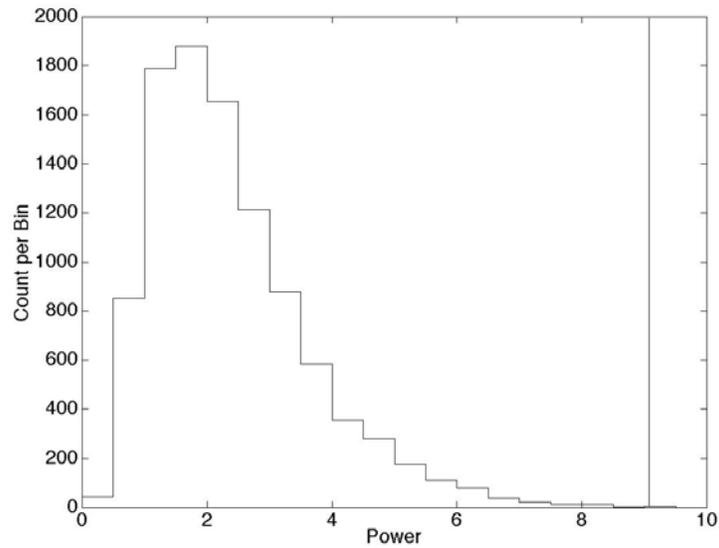

Figure 8. Histogram display of the maximum power, computed by the likelihood method, over the frequency band $39.38 - 40.54 \text{ yr}^{-1}$, for 10 000 Monte Carlo simulations of the Super-Kamiokande five-day dataset for the interval of operation of the SNO $D_2O$ experiment. Only 4 out of 10 000 simulations have power as large as or larger than the actual maximum power (9.07 at $39.38 \text{ yr}^{-1}$).

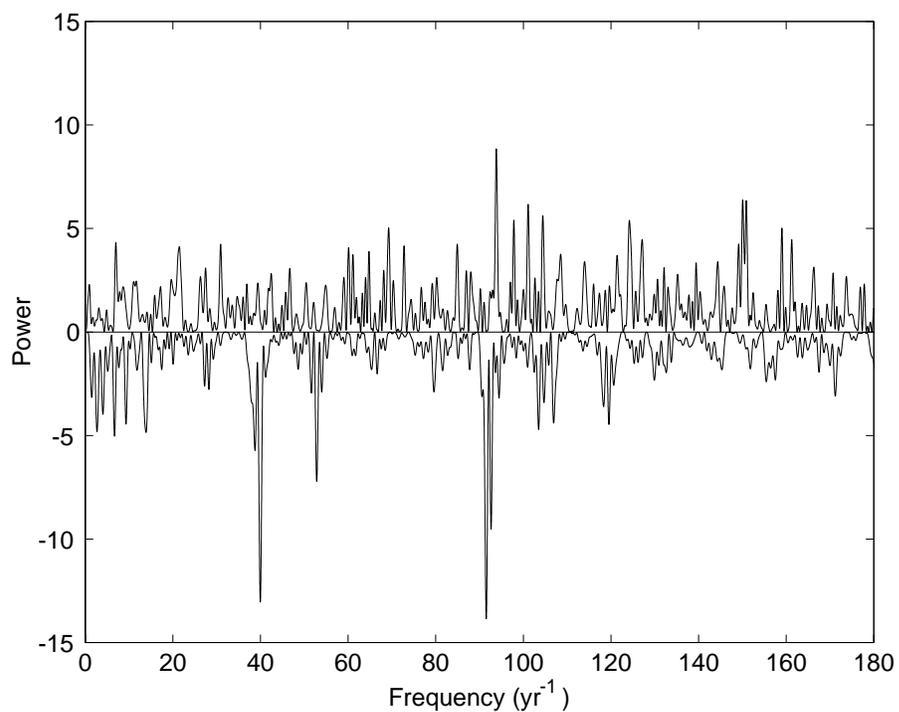

Figure 9. Display of the SNO $D_2O$ power spectrum (shown positive) and the magnetic-field power spectrum (shown negative) for the same time interval.

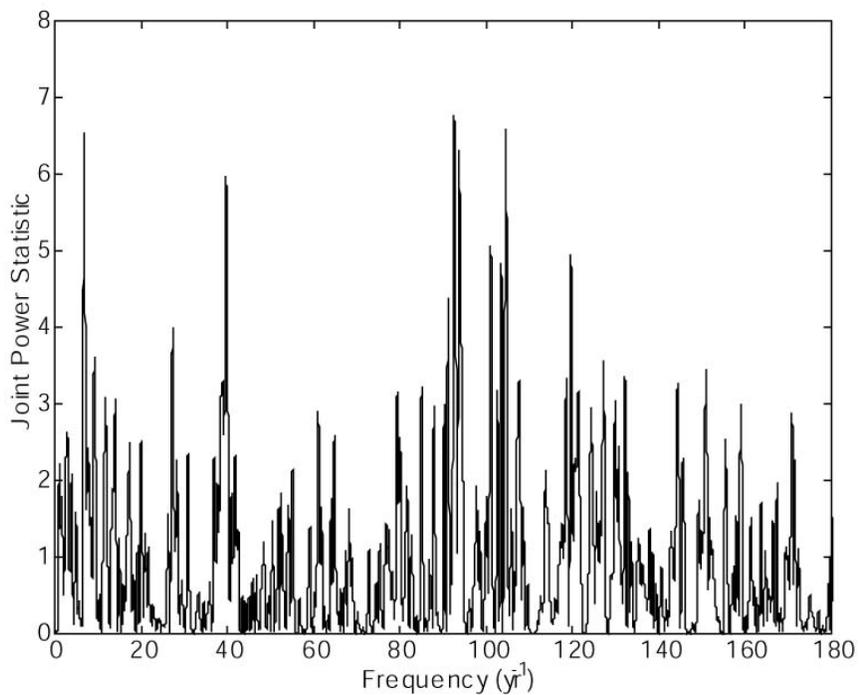

Figure 10. The joint power statistic formed from the magnetic field and SNO $D_2O$ power spectra.

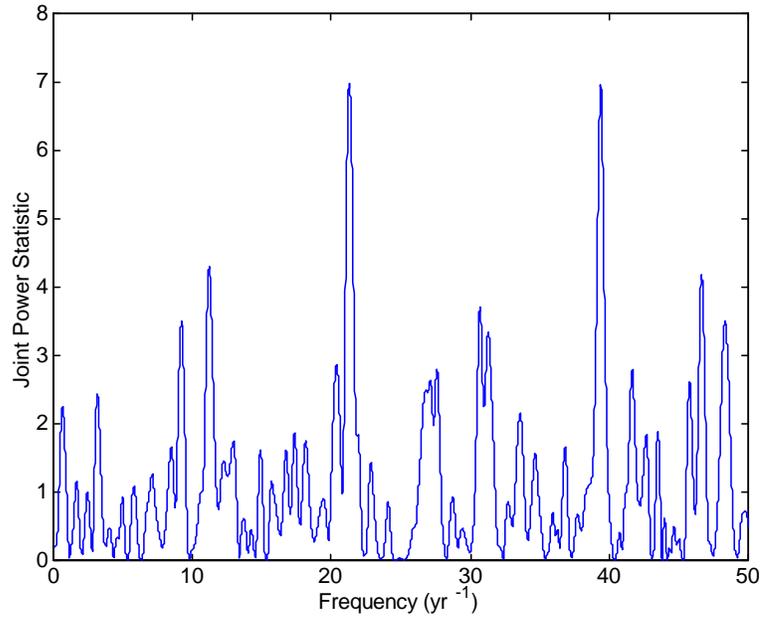

Figure 11. The joint power statistic formed from the Super-Kamiokande and SNO $D_2O$ power spectra.

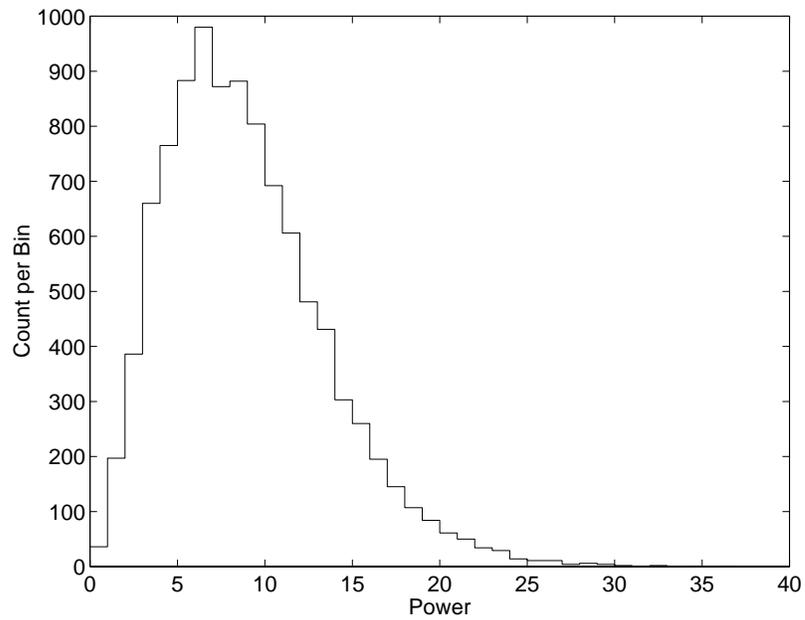

Figure 12. Histogram display of the power at $39.38\ yr^{-1}$ found from 10 000 simulations of the SNO $D_2O$ experiment for a hypothetical solar neutrino flux modulation at that frequency with 10% depth of modulation. We find that 588 simulations have power as small as or smaller than that in the actual dataset (2.43).

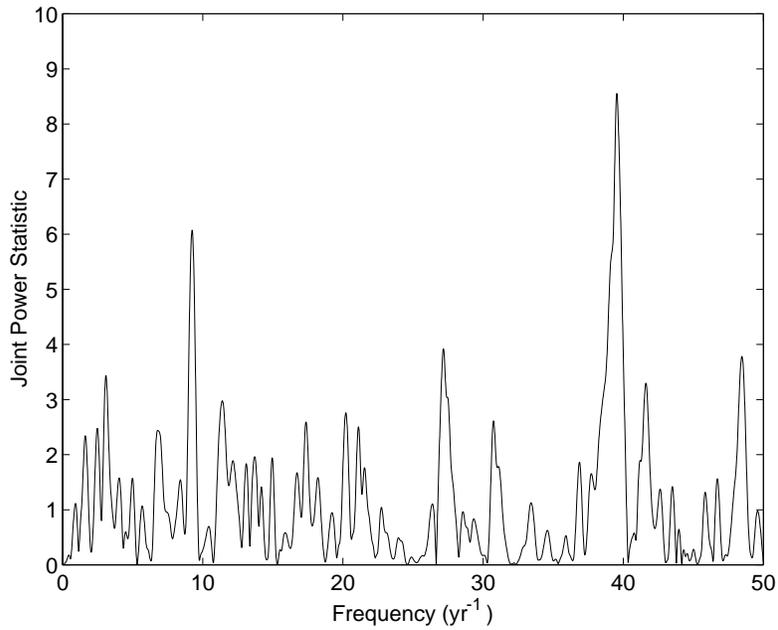

Figure 13. The joint power statistic formed from the magnetic-field, Super-Kamiokande and SNO $D_2O$ power spectra.

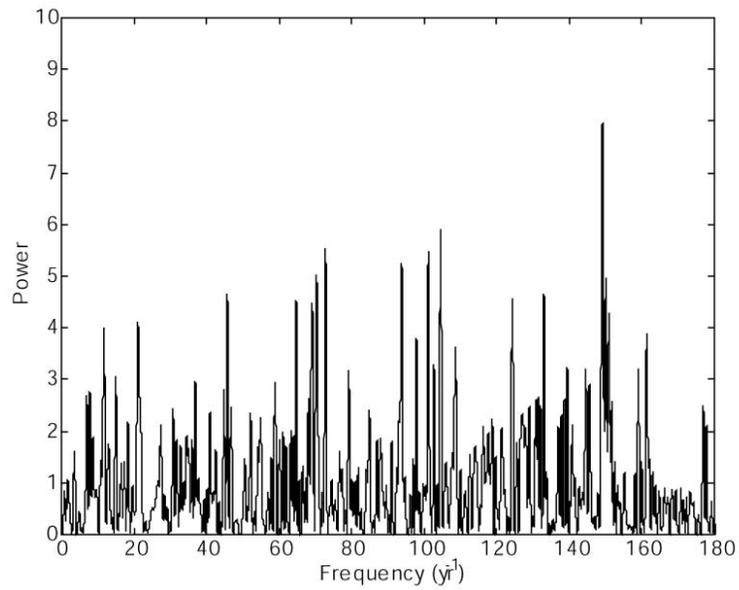

Figure 14. Power spectrum of the $D_2O$ section of the SNO dataset, computed by the likelihood procedure, but using the SNO error estimates, procedure for the frequency range $0-180\ yr^{-1}$.

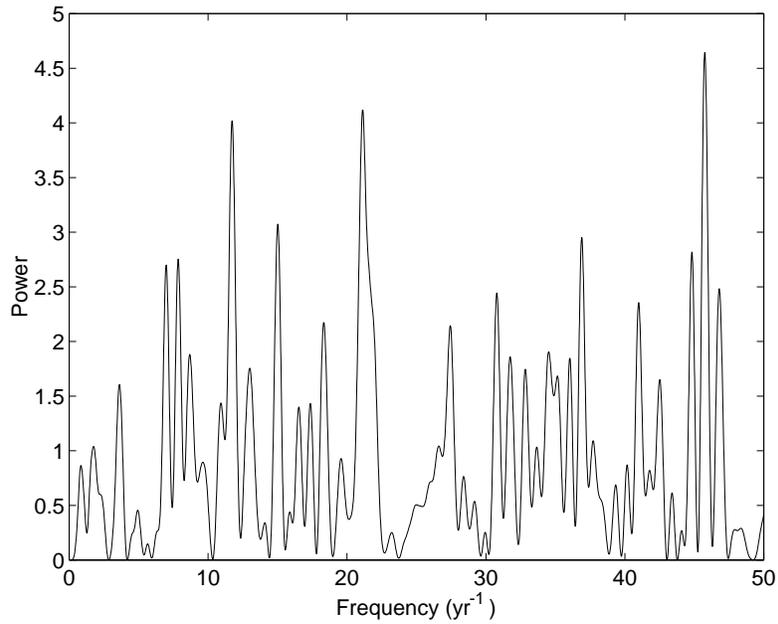

Figure 15. Power spectrum of the $D_2O$ section of the SNO dataset, computed by the likelihood procedure, but using the SNO error estimates, for the frequency range $0-50\ yr^{-1}$

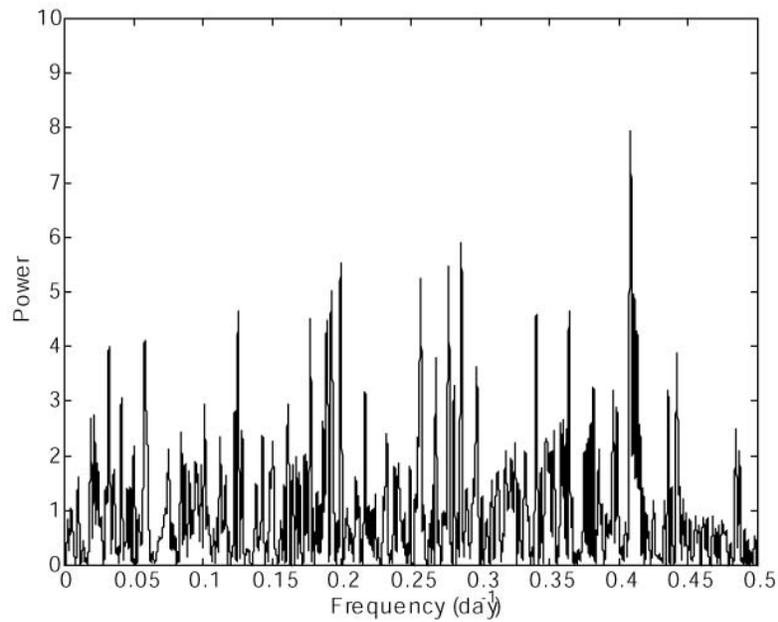

Figure 16. Power spectrum of the $D_2O$ section of the SNO dataset, computed by the likelihood procedure, but using the SNO error estimates, for the frequency range $0-0.5\ day^{-1}$.